# Absolute proof that hydrogen-antihydrogen oscillations occur in nature

G. Van Hooydonk, Ghent University, Faculty of Sciences, Krijgslaan 281, B-9000 Ghent, Belgium

**Abstract.** Detecting H-antiH oscillations is intimately connected with the existence of natural antiH. We detect natural H-antiH oscillations, generated by a classical spin-free Coulomb quantum gap, which we calculate analytically without any parameter. Oscillation times are much smaller than those predicted with the Standard Model. These unprecedented results also remove the so-called problem with matter-antimatter asymmetry in the Universe.

*Introduction*

According to recent results on the hydrogen-antihydrogen 4 particle system [1], our interest in the *charge-conjugated* Hamiltonians for *charge-symmetrical* HH and *charge-anti-symmetrical* H$\underline{H}$ systems

$$\mathbf{H}_{\pm} = \mathbf{H}_0 \pm \Delta\mathbf{H} \qquad (1)$$

led to a detailed study of long range behavior and of the role of anti-symmetry in few body systems [2]. This exercise, never done before [1], gives interesting results for physics and chemistry. We now report on an absolute proof, taken from [2], that H-$\underline{H}$ oscillations occur in nature. If a reversible H ↔ $\underline{H}$ transition is (theoretically) allowed, baryon B and lepton L numbers are not conserved and the problem is shifted towards the conservation (non-conservation, breaking…) of B-L symmetry. This is why the detection of H-$\underline{H}$ oscillations is so important [3,4].

*Theory*

Radiation appears whenever a quantum system switches states, i.e. when it goes from state A with energy $E_A$ to state B with energy $E_B$. Quantum theory relies on the simple but important relation

$$h\nu = E_A - E_B = \Delta E_{A,B} \qquad (2)$$

where $\Delta E_{A,B}$ is the *quantum gap* responsible for resonance $h\nu$. The sign of gap (2) seems irrelevant but for the physics of a system it is not: the energies of states A and B are important, including their sign with respect to a reference value like $\mathbf{H}_0$ in (1). For 4-fermion system bond $H_2$, discrete Coulomb states HH and H$\underline{H}$ appearing in (1) have different energies (symmetries) [1]. Their energy difference, *a classical Coulomb gap* $\delta$, should lead to emission/absorption of radiation, if and only if it is a quantum gap like (2). The well depth of a bond can be understood *quantitatively* with an asymptote much larger than the natural atomic threshold [1]. At longer range near atomic dissociation limit $\mathbf{H}_0$, oscillator behavior is distorted [1]. For this longer range close to the critical distance, the Heitler-London perturbation for state HH can, *arbitrarily* [1], be denoted as

$$+\Delta H = -e^2/r_{Ab} - e^2/r_{Ba} + e^2/r_{ab} + e^2/r_{AB} \tag{3a}$$

whereas the charge-conjugated *mutually exclusive* perturbation for state HH is then [1]

$$-\Delta H = -(-e^2/r_{Ab} - e^2/r_{Ba} + e^2/r_{ab} + e^2/r_{AB}) = +e^2/r_{Ab} + e^2/r_{Ba} - e^2/r_{ab} - e^2/r_{AB} \tag{3b}$$

With a 4 unit-charge model for $H_2$, relations (3) may be *exact*, but solving them is a problem [1]. At *long range*, the simplest ordered 4-particle configuration possible is of parallel dipole type ↑......↑ for *charge-symmetrical* state HH and of anti-parallel dipole type ↑......↓ for *charge-anti-symmetrical* state HH (in the arrow, the point indicates the unit charge –1, the base +1). For this biaxial model[1] to remain dynamic, the leptons must rotate in intra-atomic planes, separated by large $r_{AB}$ but perpendicular to this axis. *Without polarization*, an electrostatic description of (3) is possible, if the lepton phase difference is incorporated. This difference is zero in a *cis* model[2] but can affect the potential (see below). In this report, we skip these effects as well as many others, dealt with in extenso in [2]. With notation $R = r_{AB}$, with the Bohr radius r for intra-atomic potentials $r = r_{Aa} = r_{Bb}$ and knowing that at long range $r_{ab} = r_{AB} = R$ *in this cis form*, inter-atomic perturbations (3) simplify to

$$+\Delta H = -2e^2/\sqrt{(R^2 + r^2)} + 2e^2/R = +2(e^2/R)[1 - (1 + (r/R)^2)^{-1/2}] \tag{4a}$$

$$-\Delta H = +2e^2/\sqrt{(R^2 + r^2)} - 2e^2/R = -2(e^2/R)[1 - (1 + (r/R)^2)^{-1/2}] \tag{4b}$$

Scaling with hartree[3] $e^2/r$ and using scaled variable $m = r/R$ gives relations of type $N_\pm(m)$ or

$$N_+(m) = +2m[1 - (1 + m^2)^{-1/2}] \tag{5a}$$

$$N_-(m) = -2m[1 - (1 + m^2)^{-1/2}] \tag{5b}$$

which should lead to the character (attractive or repulsive) of perturbations (3)-(5) in the *cis* model. This conceptually simple *cis* model is of plausible Bohr type [2], which makes its treatment parameter free and analytical. Errors for *dipole-dipole interactions* (5) will be comparable with those of Bohr theory. The difference between the *charge-conjugated* quantum states (5), Coulomb gap δ, will act as a quantum gap in the sense of (2), if it leads to radiation for *allowed transitions* (oscillations hν) between the states. The analytical form of the gap, *solely due to charge-anti-symmetry*, is generated by *classical first principles* only –Coulomb's law- for the 4-fermion or 4 unit-charge system. In its analytical form

$$\delta(m) = N_-(m) - N_+(m) = -4m[1 - (1 + m^2)^{-1/2}] \sim 1/n\alpha = h\nu/(e^2/r) \tag{6}$$

this Coulomb gap will be tested below with the observed long range behavior, assessable with the band spectrum of 4-fermion system $H_2$ [5]. This will finally show if H-H oscillations hν between states HH and HH, obeying Coulomb gap (6), occur in nature. Gap (6) is peculiar in that it is spin-free: it is a pure *Coulomb gap*, due to the ambiguity of the spin- and charge-symmetries [1]. Using the available PEC (potential energy curve) of molecule $H_2$ [5], observed level energies U(R) are scaled with the hartree[3]. The numerical relation of type u(m) is

$$u(m) = [U(R) - U(\infty)]/(e^2/r) \sim 1/n\alpha = h\nu/(e^2/r) \tag{7}$$

---

[1] This is the cis configuration used in [1] with 2 linear magnets kept in a parallel/anti-parallel alignment when brought together.
[2] These cis and trans notations for structures are common in stereochemistry [2]
[3] Scaling hν with the hartree gives $h\nu/(e^2/r) = 1/n\alpha$, where α is the fine structure constant and n de Broglie's $n = 2\pi r/\lambda$.

and is sufficiently accurate to allow the direct confrontation with (6) for the long range behavior of two neutral interacting species hydrogen. To extract data restricted to this long-range behavior for the reasons above and in [1] (see for instance the left side of its Fig. 4b), we select level energies for the 11 upper *outer turning points* in the PEC, close to natural threshold **H**$_0$, the atomic dissociation limit. These 11 points at the attractive branch of the PEC are all between 17000 and 38000 cm$^{-1}$ [5]. Their average of about 30000 cm$^{-1}$ (10$^{15}$ Hz) gives average *oscillation times* of 10$^{-15}$ sec (femtosec).
*It is important to realize that all equations above are analytical and absolutely parameter free*. If the exact Bohr radius r is used (the measure for hartree 2R$_\infty$ in Rydberg, equivalent to absolute r=0,5291 Å), our calculations get *ab initio* status, despite the analytical simplicity of (6) and despite the simplicity of the geometrical model (dipole-dipole interactions, obeying the classical Coulomb law).

*Results and discussion*

- Character of *charge-symmetrical* and *charge-anti-symmetrical* perturbations (3): attractive or repulsive

Results for states (5) in a *cis* mode are in Fig. 1. With the Coulomb classification (+ is repulsive, - is attractive), N$_+$ (5a) or perturbation (3a) for *charge-symmetrical cis* HH is indeed repulsive as shown in [1] using different arguments (but see [2]). Mutually exclusive N$_-$ (5b) or *conjugated* perturbation (3b) for *charge-anti-symmetrical cis* HH is always attractive. This confirms our analysis in [1].
Despite the absolute neutrality constraint for neutral 4 unit-charge systems, it is nevertheless possible to generate a *Coulomb operator* like in (1) to distinguish *between 2 different absolutely neutral states*, one being *internally charge-symmetric*, the other *internally charge-anti-symmetric*. Because of Coulomb's law, this result in (1) is *absolute and exact*. However, for 2 unit-charge systems, this generic possibility is *excluded*, since +- ≡ -+ for the same charge separation. With only 2 fermions and Coulomb's law in Bohr atom theory, *antisymmetry can never provoke Coulomb splitting* like it does in (1), as argued in [1].
The character of HH and HH is reversed in the *trans* mode, with a maximum lepton phase difference of 180° or π rad. Here, Coulomb repulsion +e$^2$/r$_{ab}$ between the *charge-symmetrical leptons*, is attenuated [1]: for trans, r$_{ab}$=√(R$^2$+4r$^2$), whereas r$_{ab}$=R for cis. For charge-anti-symmetrical states, the extension is straightforward. The complete analysis shows that the absolute values of the potentials are always larger in *cis* than in *trans*, since *biaxial trans* is *asymmetric* but not *anti-symmetric* (full details are in [2]).

- Detection of natural H-H oscillations in the stable hydrogen molecule

An algebraic sign for both (6) and (7) gives the position of predicted (6) and observed (7) levels relative to natural threshold **H**$_0$. For (7) this means that the threshold is fixed at u(m) = 0 and that all 11 levels considered are at the *negative* (attractive) side of this threshold.

The results *for long range behavior in molecule hydrogen* are in Fig. 2, where δ(m) (6) is plotted versus u(m) (7). A line fit

$$\delta(m) = 1{,}0667 u(m) - 0{,}0103 \tag{8}$$

has a goodness of fit $R^2 = 0{,}995$. This *absolute ab initio result* is completely in line with *quantum theory* (2) for *Coulomb* gap (6). To the best of my knowledge, (8) is an unprecedented new result, which throws a completely new light on the mechanism responsible for chemical bonding, especially in prototype molecule $H_2$, completely in line with our earlier results [1].

In fact, observed long-range behavior (8) needs an even distribution of the 4 elementary particles as 4 = 2+2 [1]. This *Atoms in Molecules* (AIM) approach is gradually replaced with an odd distribution 4 =3+1, an *Ions in Molecules* (IIM) approach at closer range due to the Kratzer-potential, as argued earlier [1,6]. The transition between 2+2 and 3+1 configurations at intermediate baryon separation leads to a smooth inflection point in the *attractive branch* of the PEC, not in its *repulsive branch* [1]. But unexpected result (8) also shows that the long-range forces behind all molecular band spectra are *atom-antiatom oscillations* generated by classical Coulomb quantum gap (6), *deriving solely from charge-anti-symmetry* [1]. These oscillations obey[3] de Broglie's standing wave equation (7), conform (2) and are confined to the region of femtochemistry (see below). These and many other consequences of anti-symmetry for chemistry are in [2]. Here, we suffice by remarking that

(i)     the energy of the lowest level in Fig. 2 is 17420 cm$^{-1}$. This shows that the validity of our simple model, intended solely *for long-range behavior*, is even better in reality as it applies to about half the well depth (38000 cm$^{-1}$) and that

(ii)    the intercept in (8) shows that, at very large $r_{AB}$ ($1/r_{AB} \approx 0$), the 4-fermion system is not yet *completely ordered* as in the discrete *cis* model we used throughout at all the larger $r_{AB}$-values (due, in the first place, to *disordering* of the *cis* arrangement by thermal agitation [2]).

*Ab initio* result (8) is at least as remarkable and unprecedented for physics as it is for chemistry, since we detect a direct signature for *non-annihilation* of neutral matter H and neutral antimatter H̲, the same unconventional viewpoint as in [1]. This fits in evidence for complex annihilation patterns of heavier particle pairs [8], in clear contrast with the simple annihilation process for the electron-positron pair. First, we not only prove that natural H-H̲ oscillations really occur (8) but also that their oscillation times in a *natural* 4-fermion conglomerate are of order $10^{-15}$ sec, exactly the femtochemistry region where *longer range chemical interactions*, under study here, take place [9]. This contrasts with SM *estimates* for H-H̲ oscillations of $10^{20}$ sec [3,4]. The ratio of SM H-H̲ oscillation times and those occurring in nature, as proved by (8), is an incredibly large number $10^{20}/10^{-15} = 10^{35}$! This can never be due to a *small* error in one or more of the many parameters in the Standard Model [10,11]. We suspect that the larger part of this enormous divergence stems from the fact that *natural* H-H̲ oscillations are *embedded* within a natural and stable *4-fermion* system $H_2$=HH̲, a possibility never considered previously [1]. Moreover, it is known for long that difficulties with *4-fermion* theories were, amongst others, at the origin of the SM [12]. But the *ab initio* status of important *parameter free* result (8) for 4-fermion systems can never be overlooked. If H̲ is really an ultimate test for the Standard Model as claimed frequently, the SM has certainly not withstood the *absolute test* with the detected natural H-H̲ oscillations reported here. If confirmed, result (8) seems to be the first one, which directly flaws the highly praised but over-parameterized Standard Model [10-11].

Second, with respect to matter-antimatter asymmetry [13], we proved earlier that matter (H) is different from antimatter (H̲) [1,14]. Since our bold and unprecedented thesis [1,6] that molecule $H_2$ = HH̲ is not only confirmed but even proved with (8), the amounts of matter (H) and antimatter (H̲) *must be exactly equal on purely classical stochiometric grounds*. Since *hydrogen* is the most abundant species in the Universe, this *great cosmological question* [13] is answered, as argued in [1].

Working with Hamiltonians without wave functions, which is working with *classical physics* [1], is remarkably effective for *ab initio* calculations on difficult *quantum problems*. Signature (8) is also surprising, in that it is obtained without spin and its symmetries [1].

*Conclusion*

The present *absolute proof that natural H-H̲ oscillations exist*, combined with the results in [1], confirms that the dogmatic ban on *natural H̲* in theoretical physics and chemistry will have to be lifted as soon as possible [1].

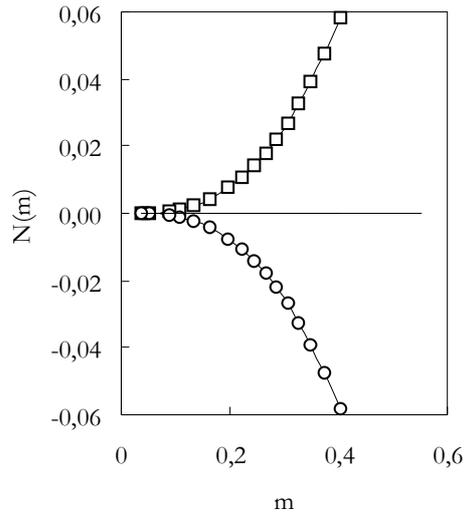

Fig. 1 N₊ for HH (squares) and N₋ for HH (circles) versus m, equations (5)

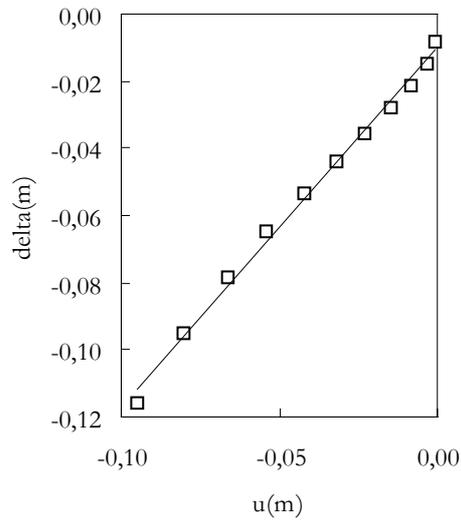

Fig. Plot of *theoretical* Coulomb gap $\delta(m)$ versus *observed* H-H oscillations u(m)